\documentclass[floatfix,aps,amsmath,nofootinbib,onecolumn,10pt]{revtex4}

\usepackage{listings}
\usepackage{graphicx}
\usepackage{bm}
\usepackage{rotating}
\usepackage{array}
\usepackage{amsmath}
\usepackage{amssymb} 
\usepackage{mathrsfs} 
\usepackage{cancel}

\def\({\left(}
\def\){\right)}
\def\[{\left[}
\def\]{\right]}

\def\e{\begin{equation}}
\def\q{\end{equation}}
\def\m{\begin{eqnarray}}
\def\n{\end{eqnarray}}

\begin{document}

\title{Parameterized tests of general relativity with gravitational wave generation and propagation}

\author{Ke Wang$^{1}$ \footnote{wangkey@lzu.edu.cn}}
\affiliation{$^1$ Institute of Theoretical Physics \& Research Center of Gravitation,\\ Lanzhou University, Lanzhou 730000, China
}

\date{\today}

\begin{abstract}
Any modification on gravity would affect not only gravitational wave (GW) generation but also GW propagation. Therefore, tests of general relativity (GR) with only GW generation or GW propagation will lead to an overestimate for deviations. Here we try to use one set of parameters to parameterize the modifications on both GW generation and GW propagation and then test GR with GW150914. In our simplest case, we find that graviton mass $\mu<6.3\times10^{-23}{\rm eV/c^2}$ at $90\%$ C.L. and there are no deviations from GR at $90\%$ C.L..

\end{abstract}

\pacs{???}

\maketitle


\section{Introduction}
\label{sec:intro}
With the detection of gravitational waves (GWs) from binary black hole (BBH) coalescence events~\cite{Abbott:2016blz,Abbott:2016nmj,TheLIGOScientific:2016pea,Abbott:2017vtc,Abbott:2017gyy,Abbott:2017oio,LIGOScientific:2018mvr}, the gravitational-wave astronomy began. After that people can test directly the Einstein's theory of gravity, general relativity (GR), on the dynamical and strong field regime with GW signals generated by BBH coalescence. Two simplest methods are the inspiral-merger-ringdown (IMR) consistency test in GR~\cite{TheLIGOScientific:2016src,Ghosh:2017gfp,LIGOScientific:2019fpa} and  constraining the parameterized deviations from IMR waveform~\cite{TheLIGOScientific:2016src,Meidam:2017dgf,LIGOScientific:2019fpa} or from ringdown waveform~\cite{Carullo:2018sfu,Isi:2019aib} in GR. 
Furthermore, without a priori assumption that GR is the correct theory, \cite{Yunes:2009ke}~build a parameterized post-Einsteinian (PPE) framework to represent the general waveform of BBH coalescence with some PPE parameters. Base on the PPE frameworks, \cite{Yagi:2012vf,Yagi:2012xdc,Hansen:2014ewa,Tahura:2018zuq,Zhang:2019iim,Carson:2019fxr,Yunes:2016jcc}~can test specific gravities on the dynamical and strong field regime.

Besides the PPE framework parameterizing the GW generation, there is a generalized GW propagation (GGP) framework~\cite{Nishizawa:2017nef} to test gravity with a general formulation of GW propagation. Unlike the PPE parameters, every GGP parameter (or function) has its own physical explanation such as propagation speed, graviton mass and a source term. During the GW propagation, these modifications on gravity will be accumulated in a particular way. Therefore, provided that GGP parameters (or functions) are obtained in specific modified gravity theories, one can constrain these gravities with GW data, as one did in~\cite{Arai:2017hxj,Wang:2020pgu,Zhao:2019xmm}.

Obviously, both of the PPE framework and the GGP framework are not a complete framework. While the former one confines itself to GW generation, the latter one confines itself to GW propagation. Undoubtedly, any modification on gravity will affect not only GW generation but also GW propagation. So, constraints on the modifications with only GW generation or GW propagation will be overestimated. Here we try to find some phenomenological relations between the PPE framework and the GGP framework and use one set of parameters to parameterize the modifications on GW generation and GW propagation at the same time, hence we can test GR properly with GW data.  

This paper is organized as follows.
In section~\ref{sec:3p}, we give a brief review of three parameterized framework for GW and try to find some phenomenological relations among them.
In section~\ref{sec:results}, we test GR with GW140914 and give the constraints on the deviations from GR.
Finally, a brief summary and discussion are included in section \ref{sec:sum}.
We adopt geometric units $c=G=1$.

\section{Three parameterized frameworks for GW}
\label{sec:3p}
The intrinsic parameters of BBH system are the BH masses $m_1$ and $m_2$ (or the mass ratio $q=m_1/m_2\geq1$ and the chirp mass $\mathcal{M}={(m_1m_2)^{3/5}}/{(m_1+m_2)^{1/5}}$ or the total mass $M=m_1+m_2$ and the symmetric mass ratio $\eta=m_1m_2/M^2$) and the dimensionless spin parameters $\chi_j=\vec{S_j}\cdot\hat{L}/m_j^2$, where the BH spin angular momenta $\vec{S_j}$ are parallel to the orbital angular momentum $\hat{L}$ and $\chi_j\in[-1,1]$. Usually, the spin effect on the GW waveform can be parameterized by a single effective spin parameter $\chi_{\rm eff}=\frac{m_1\chi_1+m_2\chi_2}{M}$ or $\chi_{\rm PN}=\chi_{\rm eff}-\frac{38\eta}{113}(\chi_1+\chi_2)$. The signal $\tilde{h}_{22}$ generated by the BBH system is given by
\begin{equation}
\tilde{h}_{22}(f;M,\eta,\chi_1,\chi_2)=A(f;M,\eta,\chi_1,\chi_2){\rm e}^{-i\phi(f;M,\eta,\chi_1,\chi_2)}.
\end{equation}
\subsection{PhenomD model}
\label{sec:D}
PhenomD model~\cite{Khan:2015jqa} divides the waveforms into three frequency regions: inspiral region with $Mf<f_1$, intermediate region with $f_1<Mf<f_2$ and merger-ringdown region with $Mf>f_2$. Then the full IMR phase is given by
\begin{equation}
\Phi_{\rm IMR}(f)=\phi_{\rm Ins}(f)\theta_{f_1=0.018}^- + \theta_{f_1=0.018}^+ \phi_{\rm Int}(f)\theta_{f_2=0.5f_{\rm RD}}^-+\theta_{f_2=0.5f_{\rm RD}}^+ \phi_{\rm MR}(f),
\end{equation}
and the full IMR amplitude is given by
\begin{equation}
\mathcal{A}_{\rm IMR}(f)=A_{\rm Ins}(f)\theta_{f_1=0.014}^- + \theta_{f_1=0.014}^+ A_{\rm Int}(f)\theta_{f_2=f_{\rm peak}}^-+\theta_{f_2=f_{\rm peak}}^+ A_{\rm MR}(f),
\end{equation}
where $f_1$ and $f_2$ are the dimensionless transition frequencies, $\theta_{f}^{\pm}$ is a step function which makes sure all three regions are joined by $C(1)$-continuous conditions, and $\phi(f)$s and $A(f)$s are polynomials in $Mf$. For example, $\phi_{\rm Ins}(f)$ and $A_{\rm Ins}(f)$ are given by
\begin{eqnarray}
\nonumber
\phi_{\rm Ins}(f)&=&2\pi f t_c -\varphi_c -\frac{\pi}{4}+\frac{3}{128\eta}(\pi Mf)^{-5/3}\sum_{j=0}^7\varphi_j(\pi Mf)^{j/3}\\
&+&\frac{1}{\eta}\left(\sigma_0+\sigma_1 Mf+\frac{3}{4}\sigma_2 (Mf)^{4/3}+ \frac{3}{5}\sigma_3 (Mf)^{5/3}+  \frac{1}{2}\sigma_4 (Mf)^{2}\right)
\end{eqnarray}
and 
\begin{equation}
A_{\rm Ins}(f)=A_0\sum_{j=0}^6\mathcal{A}_j(\pi Mf)^{j/3}+A_0\sum_{j=0}^3\rho_j( Mf)^{(6+j)/3}
\end{equation}
respectively, where $t_c$ is the coalescence time, $\varphi_c$ is the coalescence phase, $A_0=\sqrt{\frac{2\eta}{3\pi^{1/3}}}f^{-7/6}$ is a normalization factor, the PN coefficients $\varphi_j$ and $\mathcal{A}_j$ are parameterized by four physical parameters ($\eta, (m_1-m_2)/M, (\chi_1+\chi_2)/2, (\chi_1-\chi_2)/2$) and the phenomenological coefficients of higher-order terms $\sigma_j$ and $\rho_j$ are parameterized by two physical parameters ($\eta,\chi_{\rm PN}$) and calibrated against SEOBv2+NR hybrids.
\subsection{Parameterized Post-Einsteinian framework}
\label{sec:PPE}
For model-independent tests of GR, the PPN framework is introduced through the weak-field expansion of the metric tensor. Unlike the PPN, the PPE framework can test the dynamical and strong-field regime of GR through parameterizing the GW response function directly. If gravity in the dynamical and strong-field regime differs from GR, the frequency-domain PPE waveform with several PPE parameters is given by~\cite{Yunes:2009ke}
\begin{eqnarray}\tilde{h}_{22}(f)=
\begin{cases}
\tilde{h}_{{\rm Ins},22}^{\rm GR,g}(f)\cdot\left(1+\sum_j\alpha_ju^{j/3}\right){\rm e}^{i\sum_j\beta_ju^{j/3}},~~f<f_1;\\
\gamma u^c {\rm e}^{i(\delta+\epsilon u)},~~f_1<f<f_2;\\
\zeta\frac{\tau_{\rm RD}}{1+4\pi^2\tau_{\rm RD}^2\kappa(f-f_{\rm RD})^d},~~f>f_2,\\
\end{cases}
\end{eqnarray}
where $\tilde{h}_{{\rm Ins},22}^{\rm GR,g}$ is the inspiral waveform generated in GR, $u=\pi \mathcal{M}f$ is the dimensionless frequency, $f_1$ and $f_2$ are the transition frequencies, $(\alpha, a, \beta, b)$, $(c, \epsilon)$ and $(\kappa,d)$ are the PPE parameters for inspiral, merger and ringdwon respectively, $(\gamma, \delta)$ are the merger coefficients set by continuity, $\zeta$ is the ringdwon coefficient also set by continuity and $(f_{\rm RD}, \tau_{\rm RD})$ are the dominant quasi-normal (QN) frequency and decay time for ringdown. Especially, GR's prediction is obtained with $(\alpha_i, \beta_i)=(0, 0)$, $(c, \epsilon)=(-2/3,1)$ and $(\kappa,d)=(1,2)$.
\subsection{Generalized GW propagation framework}
\label{sec:GGP}
In the cosmological background, GW propagation in an effective field theory can be derived from the equation of motion of tensor perturbations~\cite{Nishizawa:2017nef}
\begin{equation}
h''_{ij}+(2+\nu)\mathcal{H}h'_{ij}+(c^2_{\rm T}k^2+a^2\mu^2)h_{ij}=a^2\Gamma\gamma_{ij},
\end{equation}
where the prime is a derivative with respect to conformal time, $a$ is the scale factor, $\mathcal{H}$ is the Hubble parameter in conformal time, $\nu$ is the Planck mass run rate, $c_{\rm T}=1-\delta_g$ is the GW propagation speed, $\mu$ is graviton mass and $\Gamma\gamma_{ij}$ is the source term from anisotropic stress. For the Friedmann-Lemaitre-Robertson-Walker (FLRW) background, the equation of motion is reduced to one with $c_{\rm T}=1$ and $\nu=\mu=\Gamma=0$. For the other  backgrounds based on a modified gravity, however, these modification terms in general are a function of time $\tau$ and wavenumber $k$. With $\Gamma=0$, the WKB solution  to the equation of motion is given by~\cite{Nishizawa:2017nef} 
\begin{eqnarray}
h&=&{\rm e}^{-\mathcal{D}}{\rm e}^{-ik\Delta T}h^{\rm GR,p},\\
\mathcal{D}&=&\frac{1}{2}\int^{\tau}\nu\mathcal{H}d\tau'=\frac{1}{2}\int_0^z\frac{\nu}{1+z'}dz',\\
\Delta T&=&\int^{\tau}\left(\delta_g-\frac{a^2\mu^2}{2k^2}\right)d\tau'=\int^z_0\frac{1}{\mathcal{H}}\left(\frac{\delta_g}{1+z'}-\frac{\mu^2}{2k^2(1+z')^3}\right)dz',
\end{eqnarray}
where $\mathcal{D}$ is the damping factor, $\Delta T$ is the time delay and $h^{\rm GR,p}$ is the time-domain monochromatic waveform which satisfies the equation of motion of tensor perturbations in GR but does not have to be generated in GR.
When all modification terms are constants and $\Gamma=0$, the WKB solution is simplified as
\begin{eqnarray}
h&=&(1+z)^{-\nu/2}{\rm e}^{-ik\Delta T}h^{\rm GR,p},\\
\Delta T&=&\frac{\delta_gd_L}{1+z}-\frac{\mu^2}{2k^2}\int^z_0\frac{dz'}{(1+z')^3\mathcal{H}},\\
d_L(z)&=&(1+z)\int^z_0\frac{dz'}{(1+z')\mathcal{H}}.
\end{eqnarray}
\subsection{Phenomenological relations among three parameterized frameworks}
\label{sec:relations}
Any modifications on gravity affects not only GW generation but also GW propagation. For GW generation, there will be some deviations from the calibrated phenomenological coefficients of PhenomD model and the PPE parameters of PPE framework will deviate from the GR's prediction. As GW propagating in the modified background, additional damping factor and time delay appear in GGP framework. Undoubtedly, changes in these three framework due to modifications on gravity should relate to each other.

The deviations from GR in PPE framework should be accumulated (or integrated) little by little during GW propagation and result in the damping factor and time delay appearing in GGP framework. By comparing waveforms in these two frameworks, for inspiral, we have
\begin{eqnarray}
\sum_j\alpha_ju^{j/3}&\stackrel{\rm int}{\longrightarrow}&-\frac{1}{2}\int_0^z\frac{\nu}{1+z'}dz',\\
\sum_j\beta_ju^{j/3}&\stackrel{\rm int}{\longrightarrow}&-k\int^z_0\frac{1}{\mathcal{H}}\left(\frac{\delta_g}{1+z'}-\frac{\mu^2}{2k^2(1+z')^3}\right)dz'.
\end{eqnarray}
If $\delta_g$, $\nu$ and $\mu$ are constants and independent on $k$, we have
\begin{eqnarray}
-2\alpha_0&\stackrel{\rm int}{\longrightarrow}&\nu\int_0^z\frac{1}{1+z'}dz',\\
-\frac{1}{2}\beta_3\mathcal{M}&\stackrel{\rm int}{\longrightarrow}&\delta_g\int^z_0\frac{1}{(1+z')\mathcal{H}}dz',\\
2\frac{\beta_{-3}}{\mathcal{M}}&\stackrel{\rm int}{\longrightarrow}&\mu^2\int^z_0\frac{1}{(1+z')^3\mathcal{H}}dz'.
\end{eqnarray}
The deviations from GR in PPE framework can be considered as ``seeds". They grow up during propagation. Without ``seeds", there should be no ``plants" and the right-hand sides become zero. Of course, without propagation ($z=0$ for the upper limit of the integral), the right-hand sides are also zero, it is because ``seeds" are not ``germinated".
Therefore, we can relate PPE parameters and GGP parameters as 
\begin{eqnarray}
-2\alpha_0&=&\nu,\\
-\frac{1}{2}\beta_3\mathcal{M}\cdot 1{\rm s}^{-1}&=&\delta_g,\\
2\frac{\beta_{-3}}{\mathcal{M}}\cdot 1{\rm s}^{-1}&=&\mu^2.
\end{eqnarray}
In SI units, for example, $\delta_g=-\frac{1}{2}\beta_3\mathcal{M}Gc^{-2}\cdot 1{\rm s}^{-1}$ and $\mu^2=2\beta_{-3}\mathcal{M}^{-1}G^{-1}c^3\cdot1 {\rm s}^{-1}\cdot\hbar^2/c^4$. Since $\delta_g$ and $\mu^2$ are independent on $k$, $\beta_{\pm3}$ are different for different BBH system in a given gravity.

Since both of PhenomD model and PPE framework deal with GW generation, we can directly add PPE parameters to PhenomD model. Then the modified insprial waveform in PhenomD model is
\begin{eqnarray}
A_{\rm Ins}(f)&=&A_0(1+\alpha_0)\sum_{j=0}^6\mathcal{A}_j(\pi Mf)^{j/3}+A_0(1+\alpha_0)\sum_{j=0}^3\rho_j( Mf)^{(6+j)/3},\\
\nonumber
\phi_{\rm Ins}(f)&=&2\pi f t_c -\varphi_c -\frac{\pi}{4}+\frac{3}{128\eta}(\pi Mf)^{-5/3}\sum_{j=0}^7\varphi_j(\pi Mf)^{j/3}\\
&+&\frac{1}{\eta}\left(\sigma_0+\sigma_1 Mf+\frac{3}{4}\sigma_2 (Mf)^{4/3}+ \frac{3}{5}\sigma_3 (Mf)^{5/3}+  \frac{1}{2}\sigma_4 (Mf)^{2}\right)-\beta_3(\pi\mathcal{M}f)-\beta_{-3}(\pi\mathcal{M}f)^{-1}.
\end{eqnarray}

Since the cosmic distance $d_L$ is much larger than the wave length $\lambda$ for BBH system, we can treat ${\rm e}^{-\mathcal{D}}{\rm e}^{-ik\Delta T}$ as constants during the Fourier transform. For the simplest case where $\delta_g$, $\nu$ and $\mu$ are constants and independent on $k$, we have the waveform of inspiral $Mf<0.018$ as
\begin{equation}
\label{eq:final}
\tilde{h}_{{\rm Ins},22}={\rm e}^{-\mathcal{D}(\alpha_0)}{\rm e}^{-2\pi if\cdot\Delta T(\beta_{-3},\beta_{3})}\cdot\tilde{h}_{{\rm Ins},22}^{\rm GR,g}(f)\cdot\left(1+\alpha_0\right){\rm e}^{i(\beta_{-3}u^{-1}+\beta_3u)}.
\end{equation}
Due the $C(1)$-continuous conditions imposed on PhenomD model, the other part of IMR waveform is also modified slightly.
\section{Constraints on the PPE parameters with GW generation and propagation}
\label{sec:results} 
Given the observed data $d(t)$ and its model $h$ with a parameter set $\theta$, we can use Bayesian inference to estimate the properties a BBH system
\begin{equation}
p(\theta|d,h)=\frac{p(d|\theta,h)p(\theta|h)}{p(d|h)},
\end{equation}
where the posterior distribution $p(\theta|d,h)$ encodes the properties of sources and the likelihood $p(d|\theta,h)$ for $N$ detectors is usually defined in frequency domain as the PyCBC Inference~\cite{Biwer:2018osg}
\begin{equation}
p(d|\theta,h)=\exp\left[-\frac{1}{2}\sum_{i=1}^N\left\langle\tilde{h}_i(f,\theta)-\tilde{d}_i(f),\tilde{h}_i(f,\theta)-\tilde{d}_i(f)\right\rangle\right].
\label{eq:like}
\end{equation}
The inner product $\langle\tilde{x},\tilde{y}\rangle$ is
\begin{equation}
\langle\tilde{x}(f),\tilde{y}(f)\rangle=4\mathfrak{R}\int_0^\infty\frac{\tilde{x}^*(f)\tilde{y}(f)}{S_n(f)}df,
\end{equation}
where $S_n(f)$ is the power spectral density of one detector's noise.

Here we will test GR with Eq.~(\ref{eq:final}). 
That is to say, we should set the waveform $\tilde{h}_i(f,\theta)$ used in the likelihood~(\ref{eq:like}) as modified IMRPhenomPv2 model. The original one shares a parameterization with PhenomD model. Here we add \{$\alpha_0,\beta_{-3},\beta_{3}$\} to it by modifying LALSuite~\citep{lalsuite}. Then the final parameter set $\theta$ consists of the PPE parameters \{$\alpha_0,\beta_{-3},\beta_{3}$\}, the intrinsic parameters of the source $\{m_1, m_2, \chi_1, \chi_2, \chi_1^a, \chi_2^a, \chi_1^p, \chi_2^p, t_c, \varphi_c\}$, the location parameters $\{d_L, \alpha, \delta\}$ and the orientation parameters $\{\psi, \iota\}$. We use uniform prior distributions for the binary component masses $m_{1,2}\in[1,120]$, uniform priors for the spin magnitudes $\chi_{1,2}\in[0.0,0.99]$, uniform solid angle priors for azimuthal angle $\chi_{1,2}^a$ and polar angle $\chi_{1,2}^p$, an uniform prior for the coalescence time $t_c\in[t_s-0.1s,t_s+0.1s]$, where $t_s$ is the trigger time and an uniform angle prior for the coalescence phase of the binary $\varphi_c$; we use an uniform volume prior for $d_L$, uniform sky position priors for the binary's right ascension $\alpha$ and declination $\delta$; we use an uniform angle prior for the polarization angle $\psi$ and a sine-angle prior for the inclination angle $\iota$; since there are degeneracies between $\nu(\alpha_0)$ and $z$ and between $\delta_g(\beta_3)$ and $d_L(z)$, we consider a simple cases: \{$\alpha_0=0,~\beta_{3}=0,~\beta_{-3}$\} and use uniform prior distributions for the third one $\beta_{-3}\in[0,3\times10^{-18}]$. 

\begin{figure}[!htp]
\begin{center}
\includegraphics[scale=0.2]{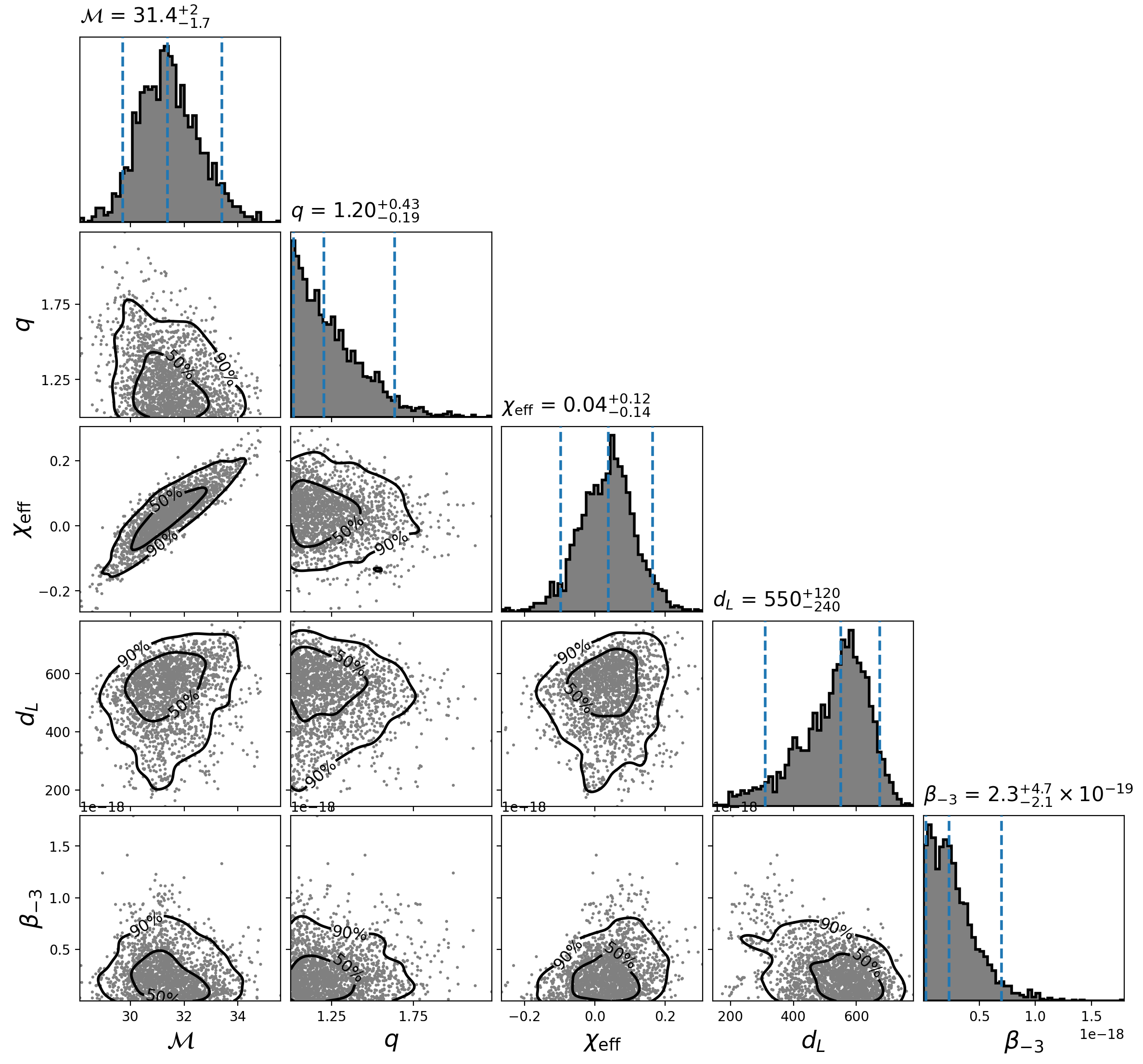}
\end{center}
\caption{The posterior distributions of the intrinsic parameters, PPE parameters and distance of GW150914 event in the detector frame. }
\label{fig:gw150914}
\end{figure}

Our final datasets have shape $\rm ntemps \times \rm nwalkers \times \rm niterations$ for the parallel-tempered sampler in PyCBC, where $\rm ntemps=20$ is the number of temperatures, $\rm nwalkers=200$ is the number of Markov chains and $\rm niterations=80000$ is the number of iterations. From them, we can plot the posterior distributions of the intrinsic parameters, PPE parameters and distance of GW150914 event in the detector frame as shown in Fig.~\ref{fig:gw150914}. We find that $\mu<6.3\times10^{-23}{\rm eV/c^2}$ at $90\%$ C.L. and there are no deviations from GR at $90\%$ C.L.. Our constraint is tighter than the former one $\mu<7.7\times10^{-23}{\rm eV/c^2}$ ~\cite{Abbott:2017vtc}. Maybe it's because the former one is overestimated. 

\section{Summary and discussion}
\label{sec:sum} 
In this paper, we first relate the PPE parameters to the GGP parameters when these GGP functions $\delta_g$, $\nu$ and $\mu$ are constants and independent on $k$. Then we use the PPE parameters \{$\alpha_0,~\beta_{-3},~\beta_{3}$\} to parameterize the modifications on GW generation and GW propagation and add them to the IMR waveform of PhenomD model. Finally, we use GW150914 to constrain \{$\beta_{-3}$\} as well as the other parameters of a BBH system. We find that there are no deviations from GR at $90\%$ C.L.. 

It is worth pointing out that the accurate relations between the PPE parameters and the GGP parameters (or functions) should derived from the explicit expressions of them in specific gravities. For example, the PPE parameters of various modified theories of gravity are listed in~\cite{Tahura:2018zuq} and the GGP functions for some gravities are listed in~\cite{Nishizawa:2017nef}. One can compare the results in the same gravity, and then relate them to each other. Of course, the final relations will be very complicated. As for the results in our paper, correctness is only not in doubt when the conditions that $\delta_g$, $\nu$ and $\mu$ are constants and independent on $k$ are satisfied.

\vspace{5mm}
\noindent {\bf Acknowledgments}

We acknowledge the use of HPC Cluster of Tianhe II in National Supercomputing Center in Guangzhou. We  would like to thank Sai Wang for his helpful discussions and advices on this paper. This research has made use of data obtained from the LIGO Open Science Center {\it https://losc.ligo.org}.



\end{document}